\documentclass[pra,floatfix,twocolumn,superscriptaddress]{revtex4-1}
\usepackage[final]{graphicx}
\usepackage{times,bbm,amsmath,amssymb}
\usepackage{epsfig,color}
\usepackage{hyperref}
\usepackage{float,siunitx}
\usepackage[caption = false]{subfig}
\usepackage[greek,english]{babel}
\usepackage{thumbpdf,enumerate}
\usepackage{booktabs}
\usepackage{sidecap}
\usepackage[scaled=.8]{couriers}
\usepackage{pstricks}
\usepackage{multirow}
\usepackage{placeins}
\usepackage{pst-grad}
\usepackage{epigraph}
\usepackage{longtable}
\usepackage{booktabs}
\usepackage{gensymb}
\usepackage{soul}
\usepackage{bbm}

\newcommand{\Tr}{{\rm Tr}}

\def\Tr{\hbox{Tr}} 
\usepackage{pifont}

\newcommand{\ket}[1]{\vert#1\rangle}
\newcommand{\bra}[1]{\langle#1\vert}

\def\ket#1{ | #1 \rangle}
\def\bra#1{{\langle #1 |  }}

\newcommand{\be}{\begin{equation}}
\newcommand{\ee}{\end{equation}}
\newcommand{\boldphi}{\boldsymbol \phi}

\begin{document}

\title{Monitoring dispersive samples with single photons: the role of frequency correlations}

\author{Emanuele Roccia}
\email{emanuele.roccia@uniroma3.it}
\affiliation{Dipartimento di Scienze, Universit\`{a} degli Studi Roma Tre, Via della Vasca Navale 84, 00146, Rome, Italy}

\author{Marco G. Genoni}
\affiliation{Quantum Technology Lab, Dipartimento di Fisica, Universit\`{a} degli Studi di Milano, 20133, Milan, Italy}

\author{Luca Mancino}
\affiliation{Dipartimento di Scienze, Universit\`{a} degli Studi Roma Tre, Via della Vasca Navale 84, 00146, Rome, Italy}

\author{Ilaria Gianani}
\affiliation{Dipartimento di Scienze, Universit\`{a} degli Studi Roma Tre, Via della Vasca Navale 84, 00146, Rome, Italy}

\author{Marco Barbieri}
\affiliation{Dipartimento di Scienze, Universit\`{a} degli Studi Roma Tre, Via della Vasca Navale 84, 00146, Rome, Italy}

\author{Marco Sbroscia}
\affiliation{Dipartimento di Scienze, Universit\`{a} degli Studi Roma Tre, Via della Vasca Navale 84, 00146, Rome, Italy}

\begin{abstract} 
The physics that governs quantum monitoring may involve other degrees of freedom than the ones initialised and controlled for probing. In this context we address the simultaneous estimation of phase and dephasing characterizing a dispersive medium, and we explore the role of frequency correlations within a photon pair generated via parametric down-conversion, when used as a probe for the medium. We derive the ultimate quantum limits on the estimation of the two parameters, by calculating the corresponding quantum Cram\'er-Rao bound; we then consider a feasible estimation scheme, based on the measurement of Stokes operators, and address its absolute performances in terms of the correlation parameters, and, more fundamentally, of the role played by correlations in the simultaneous achievability of the quantum Cram\'er-Rao bounds for each of the two parameters.
\end{abstract}

\maketitle

Quantum light provides a gentler touch when observing fragile samples \cite{arndt09,vitelli,taylor16}. While typically all the information needed can be efficiently collected through a single parameter \cite{giovannetti2011,RafalReview}, there are instances in which two parameters or more are necessary to capture the physical process under study~\cite{Paris08,Szczykulska16,Ragy17}. Such parameters might not be associated to compatible observables, hence trade-off may appear in attempts at simultaneously measuring them at the ultimate quantum precision, especially when restrictions are imposed on the resources or, in other words, to the available Hilbert space~\cite{Gill00,GenoniDispEst,Vaneph13,pinel13,Vidrighin14,knysh,safranec,Szczykulska16}.

These trade-off can be interpreted, and often circumvented, by understanding the estimation process under the geometrical standpoint by identifying the physical carrier of information with their state vectors \cite{Vidrighin14}; however, quantum probes only partly approximate such geometric entities, since these typically describe one degree of freedom at the time. The interaction with the sample might actually depend on other degrees of freedom, on which we might have limited control. A relevant example is given by dispersion effects in phase estimation: if the phase under observation depends on the optical frequency of a photonic probe, the adoption of broad bandwidths would result in dephasing~\cite{brivio10,teklu10,genoni2012,Vidrighin14}. An efficient way to tackle this is a joint estimation of the mean phase together with the characteristic width of the dephasing. 

Single photons from down-conversion are often employed as building blocks of quantum light probes~\cite{holland93,dorner2009,lee2009,datta2011,motes17}. These are produced in pairs that, under standard conditions, share frequency entanglement as a consequence of energy conservation in the generation process~\cite{kuzucu2005,nagata2007,gao2010,kacprowicz,xiang2011,jin13,harder13,wang2016,slussarenko} . In this article we calculate what impact such frequency correlation might have on the joint estimation of phase and dephasing in dispersive elements. Our study found that differences arise if one considers correlated and anti-correlated photon pairs, particularly showing how anti-correlated photons result as more interesting resources to be employed in such noisy phase estimation problem.\\

The manuscript is organized as follows: in Sec. \ref{s:lqe} we briefly review quantum estimation theory, in Sec. \ref{s:phys} we introduce our physical setting, i.e. photon pairs generated via parametric down-conversion. In Secs. \ref{s:QCRB} and \ref{s:CRB} we respectively present the ultimate limits on the estimation of phase and dephasing via correlated photons, and the performances of a feasible measurement scheme based on the measurement of Stokes operators. Sec. \ref{s:conc} concludes the paper with some final remarks and discussion.


%
\section{Multi-parameter quantum estimation theory} \label{s:lqe}
Let us consider a quantum state $\varrho_{\boldsymbol \phi}$ characterized 
by a vector of $d$ unknown parameters ${\boldsymbol \phi} = \{\phi_1,\phi_2,\dots,\phi_d\}$.
In order to estimate their value, a quantum measurement is performed, described
by POVM operators $\{ \Pi_x \}$; the whole estimation process is then purely {\em classical}, described 
by the conditional probability distribution $p(x | {\boldsymbol \phi } ) = \Tr[\varrho_{\boldsymbol \phi} \Pi_x]$
and the ultimate limits on the estimation are posed by the (multi-parameter) classical Cram\'er-Rao bound
\be
{\rm Cov} \tilde{\boldsymbol \phi} \geq \frac1M {\bf F}^{-1}(\boldsymbol \phi) \,.
\label{eq:CRB}
\ee
We have introduced the covariance of any unbiased 
estimator ${\rm Cov}\tilde \boldphi_{jk} = \mathbbm{E}[(\tilde \phi_j - \phi_j ) (\tilde\phi_k -\phi_k)]$,
where $\mathbbm{E}[\cdot]$ denotes the average over the probability distribution $p(x | \boldphi)$, and
${\bf F}(\boldphi)$ the Fisher information (FI) matrix, whose elements are evaluated as
\be
F_{jk}(\boldphi) = \mathbbm{E}\left[\left( \partial_j \log p(x|\boldphi) \right) \left( \partial_k \log p (x|\boldphi) \right) \right]\,
\ee
with $\partial_j$ denoting the partial derivative respect to the parameter $\phi_j$. The bound above is always achievable,
as unbiased optimal estimators exist allowing to saturate the inequality (\ref{eq:CRB}) in the limit of large number of measurements $M$.\\
In the quantum setting the following more fundamental bound holds,
\be
{\rm Cov} \tilde{\boldsymbol \phi} \geq \frac1M {\bf F}^{-1}(\boldsymbol \phi) \geq \frac1M {\bf Q}^{-1}(\boldsymbol \phi) \label{eq:QCRB}
\ee
where ${\bf Q}(\boldphi)$ is the quantum Fisher information (QFI) matrix, with elements
\be
Q_{jk} (\boldphi) = \Tr[\varrho_{\boldphi} (L_j L_k + L_k L_j )] /2 \,,  \label{eq:QFImatrix}
\ee
where $\{L_j\}$ denote the symmetric logarithmic derivative (SLD) operators, implicitly defined by the equation 
$2 \partial_j \varrho_{\boldphi} = L_j \varrho_{\boldphi} + \varrho_{\boldphi} L_j$. We remark that the inequalities in (\ref{eq:QCRB}) should be understood as matrix inequalities, but they can be straightforwardly translated to standard inequalities involving only variances for each parameter as
\be
\sum_j {\rm Var} (\phi_j) \geq \frac{1}{M} \Tr[{\bf F}^{-1} (\boldphi)] \geq \frac{1}{M} \Tr[{\bf Q}^{-1}(\boldphi)]  \,.\label{eq:QCRBVar}
\ee
Moreover, the following bounds hold for each single parameter, $ {\rm Var} (\phi_j) \geq \frac{1}{M} {\bf F}^{-1}_{jj} (\boldphi) \geq \frac{1}{M} {\bf Q}_{jj}^{-1}(\boldphi)$. \\
In the single-parameter scenario the ultimate quantum bound is always achievable, i.e. the existence of POVM such that the corresponding Fisher information is $F(\phi) = Q(\phi)$, is guaranteed; in particular this POVM can be easily identified as the eigenbasis of the SLD operator $L_\phi$. On the other hand, in the multi-parameter case, the quantum Cram\'er-Rao bound may not be achievable, as optimal measurements for different parameters may correspond to non-commuting observables. A necessary and sufficient condition for simultaneous achievability of the quantum Cram\'er-Rao bound (\ref{eq:QCRB}) is formulated in terms of the following weak-commutativity condition \cite{Ragy17}
\be
\Tr[\varrho_{\boldphi} [L_j, L_k] ] = 0 \,\,\,\,\,\, \forall \, \{\phi_j,\phi_k\}  \,. \label{eq:compatibility}
\ee 
This, however could imply that the optimal estimation is obtained by performing collective measurements on $n$ copies of the input states $\varrho_{\boldphi}^{\otimes n}$ \cite{Vidrighin14,Roccia17}.\\
In order to better study the trade-off in simultaneous estimation of quantum parameters, the following figure of merit has been introduced and studied in detail \cite{Gill00,Ballester04,Vidrighin14,Roccia17}:
\begin{align}
\Upsilon = \Tr[{\bf F}(\boldphi) {\bf Q}^{-1}(\boldphi) ] \leq d \,, \label{eq:Upsilon}
\end{align}
where the upper bound is a consequence of the quantum Cram\'er-Rao inequality (\ref{eq:QCRB}). Notice that, for diagonal classical and quantum FI matrices, the quantity $\Upsilon$ can be written in the simple form $\Upsilon = \sum_j F_{jj}(\boldphi) / H_{jj}(\boldphi)$. \\
If one considers single-qubit states, and only separable measurements (that is acting separately on each copy of the input state $\varrho_{\boldphi}$), one proves that $\Upsilon \leq 1$, no matter the number of parameters to be estimated \cite{Gill00,Ballester04,Vidrighin14}. In order to violate this inequality one is then left with two possible options: either consider non-separable (entangling) measurements, as suggested above and investigated in \cite{Roccia17}, or to consider states defined in a larger Hilbert space \cite{Szczykulska16}, as, for instance, correlated two-qubit probes that we will consider in the following.
\section{The physical setting}  \label{s:phys}
The starting point of our analysis is the quantum state that describes a pair of photons generated during a parametric down-conversion process:
\be
\vert \Psi_0 \rangle = \int d\omega_1 d\omega_2\, \Phi(\omega_1,\omega_2) \vert \omega_1, D \rangle \otimes  \vert \omega_2, D \rangle
\ee
where $\Phi(\omega_1,\omega_2)$ is the spectral wavefunction, and $\ket{\omega, D}$ identifies a photon at frequency $\omega$ with diagonal polarisation. Since we are interested in monitoring a dispersive medium with both copies, we consider the case where the two photons have nearly-degenerate frequencies; following the passage in the sample, the state is transformed as:
\be
\vert \Psi \rangle = \int d\omega_1 d\omega_2\, \Phi(\omega_1,\omega_2) |\psi_1 \rangle \otimes |\psi_2\rangle \,,
\ee
where
\be
|\psi_i \rangle = \frac{e^{i h(\omega_i)} \vert \omega_i, H \rangle + e^{i v(\omega_i)} \vert \omega_i, V \rangle }{\sqrt{2}} \,,
\ee
while $h(\omega)$ and $v(\omega)$ are the phases acquired respectively by the horizontal $|H\rangle$ and vertical $|V\rangle$ components.

Since the detection is frequency-insensitive, one needs to trace out the spectral part, and only consider the polarisation subspace:
\begin{align}
\rho_{\boldphi}=& {\text Tr}_{\Omega} [\vert \Psi \rangle \langle \Psi \vert]  \nonumber\\
=& \int d\omega_1 d\omega_2 \vert \Phi(\omega_1,\omega_2) \vert^2 \varrho_1(\omega_1) \otimes \varrho_2(\omega_2) \label{eq:rhopi}
\end{align}
where
\begin{align}
\textcolor{black}{\varrho_i(\omega_i)}&=\frac{\vert H_i \rangle \langle H_i \vert + \vert V_i \rangle \langle V_i \vert + e^{i \Delta_i} \vert V_i \rangle \langle H_i \vert + e^{-i \Delta_i} \vert H_i \rangle \langle V_i \vert }{2} \nonumber \\
&=\frac{1}{2} \left( \mathbb{I}+ \cos \Delta_i \hat{\sigma}_x + \sin \Delta_i \hat{\sigma}_y  \right).
\end{align}
In the formula above we have used the shortcut notation $\vert H_i \rangle = \vert \omega_i,H \rangle$,  $\vert V_i \rangle = \vert \omega_i,V \rangle$, and $\Delta_i=\Delta(\omega_i)=v(\omega_i)-h(\omega_i)$ with $i=1,2$, \textcolor{black}{introducing the Pauli operators $\hat{\sigma}_x=\bigl(\begin{smallmatrix} 0&1\\ 1&0 \end{smallmatrix}\bigr)$, $\hat{\sigma}_y=\bigl(\begin{smallmatrix} 0&-i\\ i&0 \end{smallmatrix}\bigr)$}. We notice from Eq. (\ref{eq:rhopi}) that, by acting with the trace, we are only sensitive to classical correlations in frequency, hence oblivious of the presence of coherence. Based on this decomposition, the relevant parameter to be assessed is the phase difference $\Delta_i$, which is typically inferred by measuring the Stokes operators $\hat X_i=2\ket{D_i}\bra{D_i}-\mathbb{I}$, and $\hat Y_i=2\ket{R_i}\bra{R_i}-\mathbb{I}$ ($\ket{R}$ is the right-circular polarisation), on each photon of the pair. Hence, we can write the expectation values for the Stokes operators, for instance as:
\be
\langle \hat{X_1} \hat{X_2} \rangle = \int d\omega_1 d\omega_2 \vert \Phi(\omega_1,\omega_2) \vert^2 \cos(\Delta(\omega_1)) \cos(\Delta(\omega_2)) \,.
\ee
In this manuscript we will restrict to spectral wavefunctions of the form
\be
|\Phi(\omega_1,\omega_2)|^2=\frac{e^{-(\omega_1 - \omega_2)^2/(2 \sigma_-^2)}e^{-(\omega_1+\omega_2 - 2\omega_0)^2/(2 \sigma_+^2)}}{ \pi \sigma_+ \sigma_-}  \, \label{spectralw}
\ee
where 
\be
\sigma_\pm^2 = 2 \sigma^2 ( 1 \pm \epsilon ) \,, \,\,\,\,\,\, -1 \leq \epsilon \leq 1 \,.
\ee
Even if these functions do not constitute the most general expression for correlated photons, they are sufficient to capture the basic features of our problem while allowing for simple analytical expression.
For $\epsilon=0$ the two photons are uncorrelated, i.e. 
\be
|\Phi(\omega_1,\omega_2)|^2 = \prod_{j=1}^2 \frac{e^{-(\omega_j - \omega_0)}/(2\sigma^2)}{\sqrt{2 \pi \sigma^2}} \,,
\ee
while for $\epsilon=\pm 1$ \textcolor{black}{the spectral wavefunction in Eq. (\ref{spectralw}) converges to a Gaussian multiplied by a delta function, i.e. the photons are perfectly correlated or anti-correlated in frequency, respectively.}\\
We can now Taylor expand the phase difference $\Delta(\omega)$  around the central frequency of the photons $\omega_0$ up to first order
\be
\Delta(\omega) \approx \phi_0 + \phi_1 ( \omega - \omega_0) \,.
\ee
Here $\phi_0$ is the average value normally considered in phase estimation problems, while $\phi_1$ is the first term that appears due to dispersion in the medium, and is typically responsible for phase diffusion on the qubit state. In the following we will focus on the joint estimation of these two parameters, studying in detail the role played by the correlations between the two photons.
\section{Ultimate quantum bounds on phase and dephasing estimation with correlated pair of photons}\label{s:QCRB}
In this section we will discuss the ultimate bounds posed by the QFI matrix for the parameters $\boldphi = \{\phi_0, \phi_1\}$, considering the input state $\rho$ in Eq. (\ref{eq:rhopi}). The SLD operators $L_{\phi_0}$ and $L_{\phi_1}$ can be evaluated (at least numerically) by means of the formula \cite{Paris08}
\begin{align}
L_{\phi_j} = 2 \sum_{s,t} \frac{\langle \lambda_s | \partial_{j} \varrho_{\boldphi} | \lambda_t \rangle}{\lambda_s + \lambda_t} |\lambda_s\rangle\langle \lambda_t | \,,
\end{align}
where $\{|\lambda_s\rangle\}$ and $\{\lambda_s\}$ are eigenvectors and corresponding eigenvalues of the quantum state $\varrho_{\boldphi}$. The QFI matrix elements are straightforwardly evaluated as in Eq. (\ref{eq:QFImatrix}). We numerically obtain that the off-diagonal elements are zero, so that the diagonal elements directly quantify the ultimate precision achievable on each of the two parameters. We remind that these limits may be achieved in the single-parameter scenario, that is if the value of the other parameter characterizing the quantum state is already known.\\

\par
\begin{figure}[t!]
\includegraphics[width=\columnwidth]{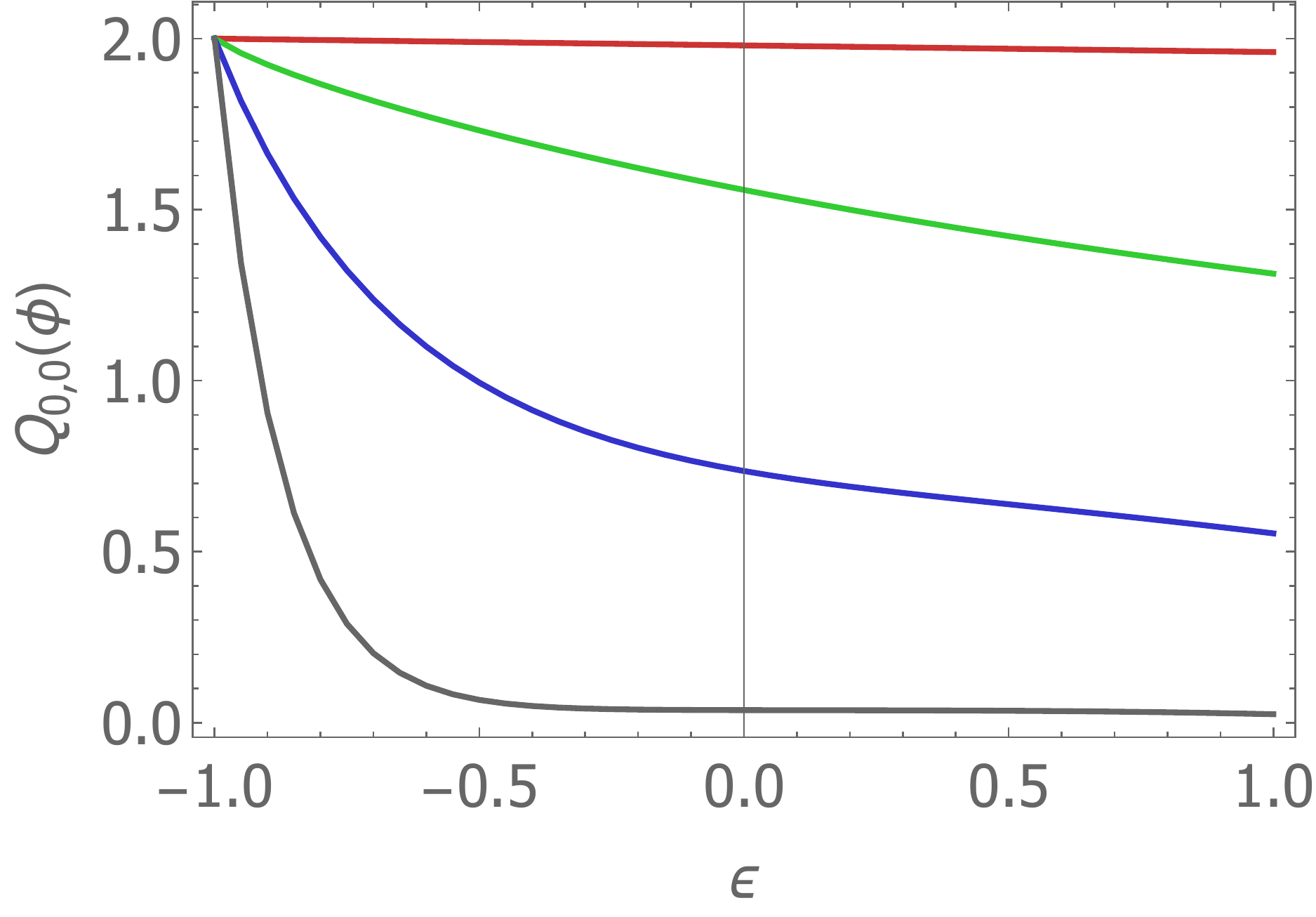}
\caption{QFI matrix element $Q_{0,0}(\boldphi)$ as a function of the correlations parameter $\epsilon$, for $\sigma^2 =1$ and for different values of dephasing. From top to bottom $\phi_1 = \{ 0.1, 0.5, 1,2\}$. \textcolor{black}{We checked numerically that $Q_{1,1}(\boldphi)$ is indipendent on the value of the phase $\phi_0$.}
\label{f:QFIphi0}}
\end{figure}

We start by focusing on the estimation of the phase $\phi_0$: in Fig. \ref{f:QFIphi0} we plot the QFI element $Q_{0,0}(\boldphi)$ as a function of $\epsilon$ for different values of $\phi_1$ (notice that we numerically checked that $Q_{0,0}$ is independent on the value of $\phi_0$). We observe how the value of the corresponding QFI is, as expected, monotonically decreasing with the dephasing parameter $\phi_1$. Remarkably we also conclude that the QFI is monotonically decreasing with $\epsilon$, that is anti-correlated photons are more sensitive to small variations of $\phi_0$, whenever some dephasing due to the dispersive medium acts on them. One also obtains that, for a maximally anti-correlated state (i.e. for $\epsilon=-1$) the QFI seems to be independent on $\phi_1$, and equal to its maximum value $Q_{0,0} = 2$.

\par
\begin{figure}[t!]
\includegraphics[width=\columnwidth]{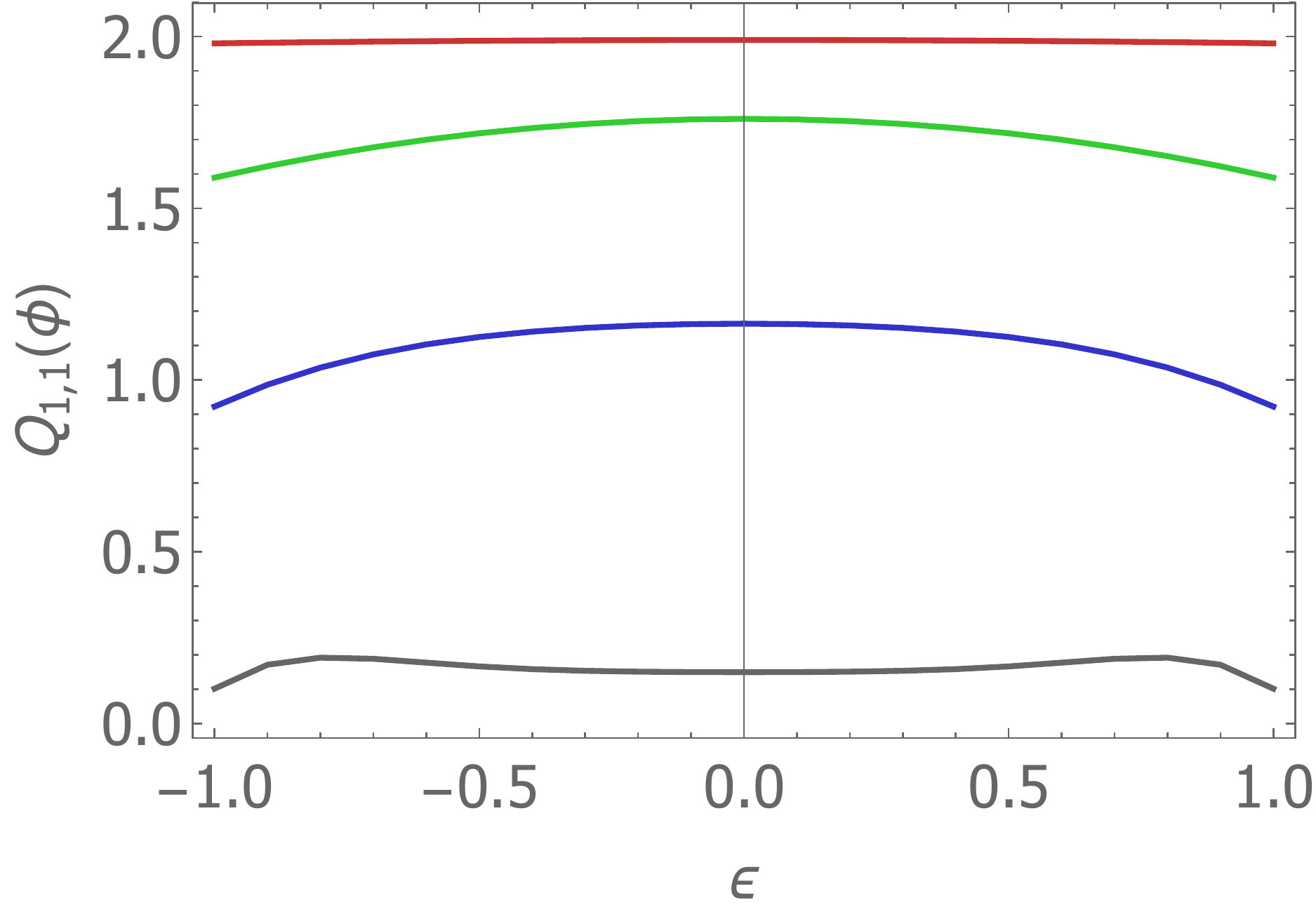}
\caption{QFI matrix element $Q_{1,1}(\boldphi)$ as a function of the correlations parameter $\epsilon$, for $\sigma^2 =1$ and for different values of dephasing. From top to bottom $\phi_1 = \{ 0.1, 0.5, 1,2\}$. \textcolor{black}{We checked numerically that $Q_{1,1}(\boldphi)$ is indipendent on the value of the phase $\phi_0$.}
\label{f:QFIphi1}}
\end{figure}

We now address the estimation of the dephasing parameter $\phi_1$ at fixed (and known) values of $\phi_0$. In Fig. \ref{f:QFIphi1} we plot as above the corresponding QFI element $Q_{1,1}$ as a function of the correlation parameter $\epsilon$ for different values of $\phi_1$. Also in this case the QFI matrix element monotonically decreases, as we expected, with the noisy parameter $\phi_1$. However the behaviour as a function of $\epsilon$ depends on the particular value of the dephasing: for small values of $\phi_1$, uncorrelated states result to be the optimal state, that is $Q_{1,1}$ is maximized for $\epsilon=0$. On the other hand, for $\phi_1$ larger than a critical value $\phi_1^*$, two maxima appears for symmetric values of $\epsilon$, showing how correlation may enhance the estimation of the parameter.
%
\section{A feasible measurement scheme: single-parameter and multi-parameter estimation} \label{s:CRB}
\begin{figure}[t!]
\includegraphics[width=\columnwidth]{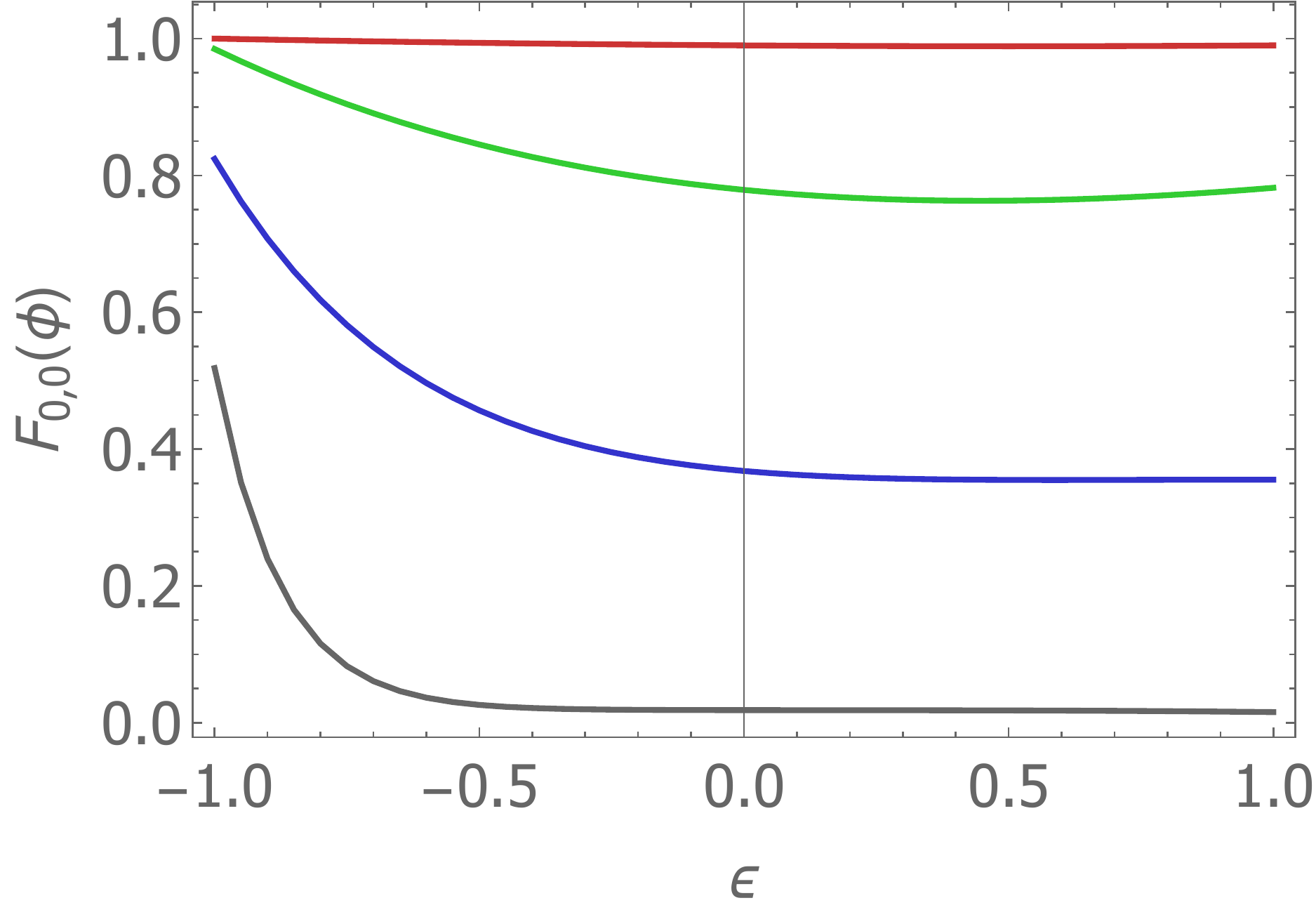}
\caption{FI matrix element $F_{0,0}(\boldphi)$ for the Stokes-operators based POVM $\{\Pi_{jk}\}$, as a function of the correlations parameter $\epsilon$, for $\sigma^2 =1$, \textcolor{black}{$\phi_0 = k \pi/4$}, and for different values of dephasing. From top to bottom $\phi_1 = \{ 0.1, 0.5, 1,2\}$.
\label{f:FIphi0}}
\end{figure}

\begin{figure}[t!]
\includegraphics[width=\columnwidth]{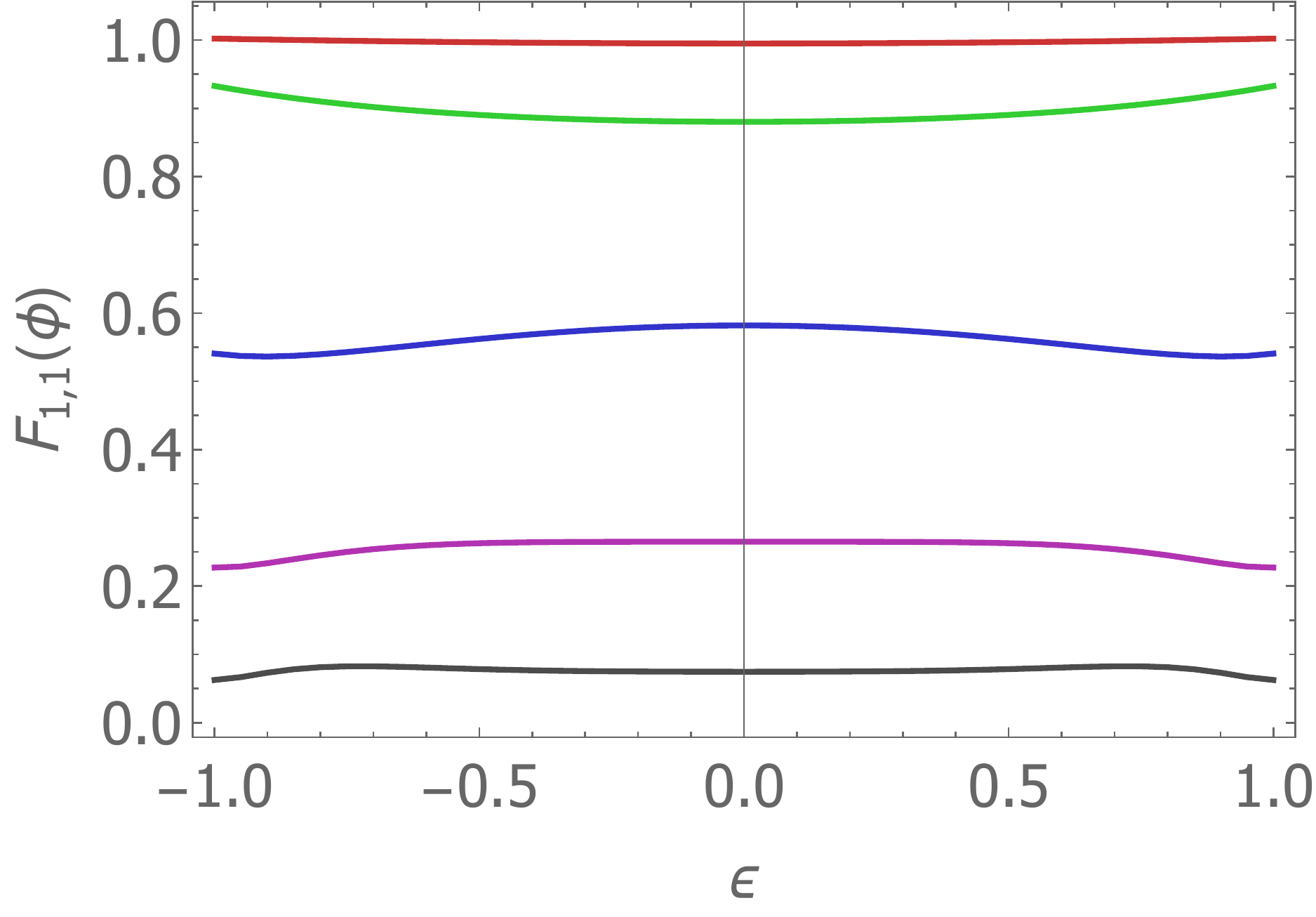}
\caption{FI matrix element $F_{1,1}(\boldphi)$ for the Stokes-operators based POVM $\{\Pi_{jk}\}$, as a function of the correlations parameter $\epsilon$, for $\sigma^2 =1$, \textcolor{black}{$\phi_0 = k \pi/4$}, and for different values of dephasing. From top to bottom $\phi_1 = \{ 0.1, 0.5, 1, 1.5, 2\}$.
\label{f:FIphi1}}
\end{figure}

We now consider a feasible estimation scheme, based on the measurement of the Stokes operators $\hat{X}$ and $\hat{Y}$ on each photon. Mathematically it is described by (multi-indexed) POVM operators $\{\Pi_{jk} = \pi_j \otimes \pi_k \}$, where 
\be
\{ \pi_j \} = \left\{ \frac{|D\rangle\langle D|}{2} , \frac{|A\rangle\langle A|}{2} , \frac{|R\rangle\langle R|}{2} , \frac{|L\rangle\langle L|}{2} \right\}
\ee
is a POVM acting on the single photon, physically corresponding to measuring half of the times $\hat{X}$ and half of the times $\hat{Y}$. From the conditional probabilities $p(jk | \boldphi) = \Tr[\varrho \Pi_{jk}]$ one can straightforwardly evaluate the corresponding classical Fisher Information matrix. In the following we will restrict ourselves to the case $\phi_0 = k \pi/4$, where we numerically obtain that the off-diagonal matrix are equal to zero. Other schemes based on the use of entangling measurements can offer in principle an advantage in the joint parameter estimation, however their usefulness is limited to the low-dephasing regime~\cite{Roccia17}.
We start by discussing, as in the previous section, the estimation of each single parameter, assuming that the other parameter is known. Given the nature of the POVM, designed in order to gain information on both $\phi_0$ and $\phi_1$, the single Cram\'er-Rao bounds will not be saturated, however we are interested in studying in more detail the role played by the correlations for this particular estimation strategy, starting from the single-parameter scenario.\\
In Fig. \ref{f:FIphi0} we plot the FI matrix element $F_{0,0}(\boldphi)$ corresponding to the estimation of the phase $\phi_0$, as a function of $\epsilon$ and for different values of dephasing. As for the QFI, we obtain that $F_{0,0}(\boldphi)$ is monotonically decreasing both with the dephasing $\phi_1$ and with the correlation parameter $\epsilon$, showing how anti-correlated photons are more sensitive to small variation of the phase, as predicted also by the QFI calculations. \\
Similarly, in Fig. \ref{f:FIphi1} we plot the FI $F_{1,1}(\boldphi)$ corresponding to the estimation of the dephasing parameter $\phi_1$; while, as we expected also in this case the Fisher information is monotonically decreasing with $\phi_1$, its behaviour is symmetric for positive or negative values of $\epsilon$ and three different regimes can be identified: for small values of $\phi_1$ optimality is achieved for maximally correlated or anti-correlated photons (i.e. for $\epsilon=\pm 1$); for intermediate dephasing, uncorrelated states are optimal, while for larger $\phi_1$ two symmetric maxima appeared for the FI, in correspondence of two values of $\epsilon = \pm \tilde{\epsilon}$, with $0 < \tilde{\epsilon} <1$.\\

\par
\begin{figure}[t!]
\includegraphics[width=\columnwidth]{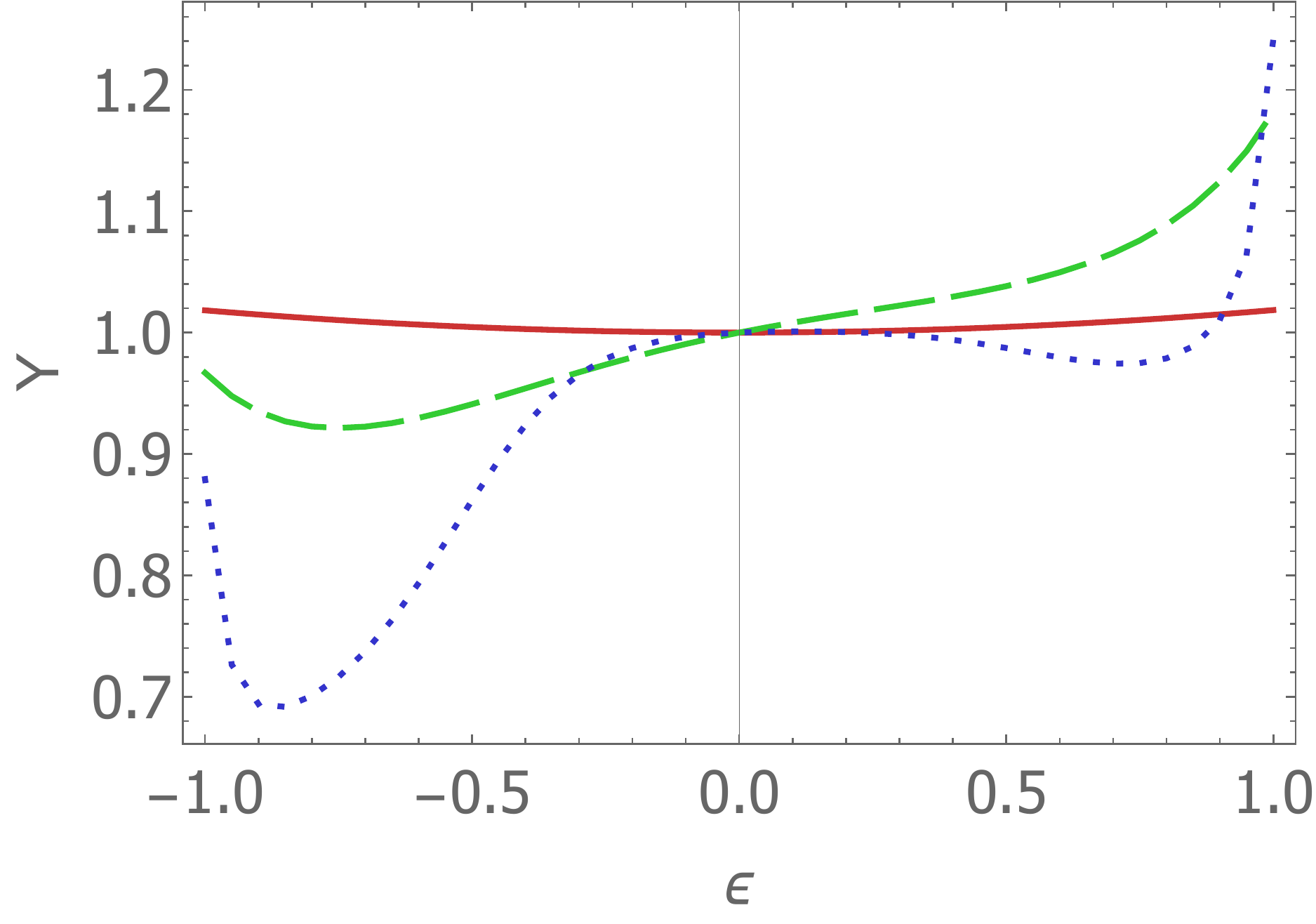}
\caption{Joint-estimation figure of merit $\Upsilon$, as a function of the correlations parameter $\epsilon$, for $\sigma^2 =1$, \textcolor{black}{$\phi_0 = k \pi/4$}, and for different values of dephasing: red-solid line, $\phi_1=0.1$; green-dashed line, $\phi_1 = 1$; blue-dotted line, $\phi_1 =2$. \label{f:Upsilon}}
\end{figure}

We now address the more fundamental problem of the joint estimation of the two parameters via the POVM $\{\Pi_{jk}\}$, studying the behaviour of the figure of merit $\Upsilon$ defined in Eq. (\ref{eq:Upsilon}). The behaviour of $\Upsilon$ as a function of $\epsilon$ for different values of $\phi_1$ is plotted in Fig. \ref{f:Upsilon}. For $\epsilon=0$ the two qubits are uncorrelated and we are left with the same situation described in \cite{Vidrighin14}, such that $\Upsilon=1$. On the other hand we observe that for correlated photons, in particular for $\epsilon >0$ it is possible to surpass the single-qubit bound, and that in particular the largest values of $\Upsilon$ are obtained for $\epsilon=1$. 

\section{Conclusion} \label{s:conc}

Phase estimation in dispersive samples is affected by frequency correlations introducing non-classical coupling. This has consequences on the precision limit achievable with the quantum metrological scheme.
The estimation strategy usually aims at finding the optimal measurement to saturates the corresponding Cram\'er-Rao bound. However, when dealing with multi-parameter scenarios, the situation is further complicated and one should study the behaviour of all the interested figures of merit (i.e. ${\bf F}$, ${\bf Q}$ and $\Upsilon$).\\
The results we have presented illustrate how both the FI matrix and the QFI matrix depend on the input state $\varrho_{\boldphi}$, and in particular on the correlation parameter $\epsilon$. Our analysis reveals that there exist cases in which maximising the estimability by maximising $\Upsilon$ does not necessarily correspond to achieving maximal information. In fact in Fig. \ref{f:Upsilon} we show that values of $\Upsilon$ larger than the single photon bound are observed for $\epsilon>0$; however, we previously showed that  the Fisher Information on the phase $\phi_0$ is improved for anti-correlated pairs, i.e. for $\epsilon<0$. 

\textcolor{black}{This increase can be related to the capabilities of dispersion cancellation in optical coherence tomography using the Hong-Ou-Mandel effect~\cite{Mazurek2013,Abouraddy2002,Strekalov1998,Nasr2003,Banaszek2007}; in these case as well, more information, in the form of improved spatial resolution can be achieved when using frequency anti-correlated photon pairs, although in an interferometric setup. Remarkably, the advantage relies on frequency anticorrelation rather than on entanglement, as recently demonstrated in experiments~\cite{Mazurek2013}.}
In general, we remark that it is typically advantageous to consider the most informative state regardless the saturability of the associated QCRBs for each single parameter.

From an alternative perspective, we can consider that frequency correlations in the initial photon pair results in correlated noise on the two photons during the estimation. The improvement we predict is consistent with the capability of correlated noisy channels to bring about non-classical features, as in~\cite{PRL111}.
In conclusion we have discussed how a physical property of photon pairs, namely their frequency correlations, should be taken into account when one studies their use for phase estimation. This example highlights how the presence of correlations in the probe states determines not only the saturability of the QCRB, but also, independently, the amount of information on the individual parameters.


\section{Acknowledgments} 
This work has been supported by the EC project QCUMbER (grant no. 665148). MGG acknowledges support from Marie Skłodowska-Curie Action H2020-MSCA-IF-2015 (project ConAQuMe, grant no. 701154).


\begin{thebibliography}{99}
\bibitem{arndt09} M. Arndt, T. Juffmann and V. Vedral, HFSP J. {3}, 386-400 (2009).
\bibitem{vitelli} F. Wolfgramm, C. Vitelli, F. A. Beduini, N. Godbout, and M. W. Mitchell, Nature Photonics {\bf7}, 28–32 (2013).
\bibitem{taylor16} M. Taylor and W. P. Bowen, Phys. Rep. {\bf 615} 1-59 (2016).
\bibitem{giovannetti2011} V. Giovannetti, S. Lloyd and L. Maccone, Nat. Photon. {\bf 5}, 222 (2011) .
\bibitem{RafalReview} R. Demkovicz-Dobrzanski, M. Jarzyna and J. Kolodynski, Progress in Optics {\bf 60}, 345 (2015).
\bibitem{Paris08} M.G.A. Paris, { Int. J. Quantum Info.} {\bf 7}, 125-137 (2009).
\bibitem{Szczykulska16} M. Szczykulska, T. Baumgratz, and A. Datta, {Adv. Phys. X} {\bf 1}, 621-639 (2016).
\bibitem{Ragy17} S. Ragy, M. Jarzyna, and R. Demkowicz-Dobrza\'nski, { Phys. Rev. A } {\bf 94}, 052108 (2016).
\bibitem{Gill00}  R. D. Gill, and S. Massar, {Phys. Rev. A} {\bf 61}, 042312 (2000).
\bibitem{GenoniDispEst} M. G. Genoni, M. G. A. Paris, G. Adesso, H. Nha, P.L. Knight and M.S. Kim, Phys. Rev. A {\bf 87}, 012107 (2013).
\bibitem{Vaneph13} C. Vaneph, T. Tufarelli and M.G. Genoni, Quantum Meas. Quantum Metr. {\bf 1}, 12 (2013).
\bibitem{pinel13} O. Pinel, P. Jian, N. Treps, C. Fabre, and D. Braun, Phys. Rev. A {\bf88} 040102(R) (2013).
\bibitem{Vidrighin14} M.D. Vidrighin, {\it et al.}, {Nat. Commun.} {\bf 5}, 3532 (2014).
\bibitem{knysh} S.I. Knysh, and G.A. Durkin arXiv:1307.0470.
\bibitem{safranec}D. {\v{S}afr\'anek}, A. R. Lee, and I. Fuentes, New J. Phys. {\bf 17} 073016 (2015).
 \bibitem{Szczykulska17} M. Szczykulska , T. Baumgratz and A. Datta, Quantum Sci. Technol. {\bf 2}, 044004 (2017).
\bibitem{brivio10} D. Brivio, S. Cialdi, S. Vezzoli, B. Teklu, M.G. Genoni, S. Olivares, and  M.G.A. Paris, Phys. Rev. A {\bf 81} 012305 (2010).
\bibitem{teklu10} B. Teklu, M.G. Genoni, S. Olivares, and M.G.A. Paris, Phys. Scr. {\bf T140} 014062 (2010).
\bibitem{genoni2012} M.G. Genoni, S. Olivares, D. Brivio, S. Cialdi, D. Cipriani, A. Santamato, S. Vezzoli, and M.G.A. Paris, Phys. Rev. A {\bf 85} 043817 (2012).
\bibitem{holland93} M. J. Holland and K. Burnett, Phys. Rev. Lett. {\bf71}, 1355 (1993). 
\bibitem{datta2011} A. Datta, L. Zhang, N. Thomas-Peter, U. Dorner, B. J. Smith, and Ian A. Walmsley, Phys. Rev. A 83, 063836 (2011).
\bibitem{dorner2009} U. Dorner, R. Demkowicz-Dobrzanski, B. J. Smith, J. S. Lundeen, W. Wasilewski, K. Banaszek, and I. A. Walmsley Phys. Rev. Lett. 102, 040403 (2009).
\bibitem{lee2009} T.-W. Lee, S. D. Huver, H. Lee, L. Kaplan, S. B. McCracken, C. Min, D. B. Uskov, C. F. Wildfeuer, G. Veronis, and J. P. Dowling, Phys. Rev. A 80, 063803 (2009).
\bibitem{motes17} K.R. Motes, J.P. Olson, E.J. Rabeaux, J.P. Dowling, S.J. Olson, and P.P. Rohde, Phys. Rev. Lett. {\bf 114} 170802 (2015).
\bibitem{kuzucu2005} O. Kuzucu, M. Fiorentino, M. Albota, F. N. C. Wong, and F. K{\"a}rtner, Phys. Rev. Lett. {\bf 94}, 083601 (2005).
\bibitem{nagata2007} T. Nagata, R. Okamoto, J. L. O’Brien, K. Sasaki, and S. Takeuchi, Science {\bf316}, 726 (2007).
\bibitem{gao2010} W.-B. Gao, C.-Y. Lu, X.-C. Yao, P. Xu, O. G{\"u}hne, A. Goebel, Y.-A. Chen, C.-Z. Peng, Z.-B. Chen, and J.-W. Pan, Nat. Phys. {\bf6}, 331 (2010).
\bibitem{kacprowicz} M. Kacprowicz, R. Demkowicz-Dobrza{\'n}ski, W. Wasilewski, K. Banaszec, and I.A. Walmsley, Nat. Photon. {\bf 4} 357-360 (2010).
\bibitem{xiang2011} G. Y. Xiang, B. L. Higgins, D. W. Berry, H. M. Wiseman and G. J. Pryde, Nat. Photon. {\bf5}, 43 (2011).
\bibitem{jin13} R.-B. Jin, R. Shimizu, K. Wakui, H. Benichi, and M. Sasaki,
 Opt. Exp. 21, 10659–10666 (2013).
\bibitem{harder13} G. Harder, V. Ansari, B. Brecht, T. Dirmeier, C. Marquardt, and C. Silberhorn, Opt. Exp. 21, 13975–13985 (2013).
\bibitem{wang2016} X.-L. Wang, L.-K. Chen, W. Li, H.-L. Huang, C. Liu, C. Chen, Y.-H. Luo, Z.-E. Su, D. Wu, Z.-D. Li, H. Lu, Y. Hu, X. Jiang, C.-Z. Peng, L. Li, N.-L. Liu, Y.-A. Chen, C.-Y. Lu, and J.-W. Pan, Phys. Rev. Lett. {\bf117}, 210502 (2016).
\bibitem{slussarenko} S. Slussarenko, M. M. Weston, H. M. Chrzanowski, L. K. Shalm, V. B. Verma, S. W. Nam and G. J. Pryde, Nat. Photon.  {\bf11}, 700 (2017).
\bibitem{Roccia17} E. Roccia, {\it et al.}, Quantum Sci. Tech. {\bf3}, 01LT01 (2018).
\bibitem{Ballester04} M. A. Ballester, Phys. Rev. A {\bf 69}, 022303 (2004);\\ M. A. Ballester Phys. Rev. A {\bf 70}, 032310 (2004).
\bibitem{PRL111}B. P. Lanyon, P. Jurcevic, C. Hempel, M. Gessner, V. Vedral, R. Blatt, and C. F. Roos, Phys. Rev. Lett. {\bf111} 100504 (2013).

\bibitem{Mazurek2013} M. D. Mazurek, K. M. Schreiter, R. Prevedel, R. Kaltenbaek and K. J. Resch, Sci. Rep. {\bf 3}, 1 (2013).
\bibitem{Abouraddy2002} A.F. Abouraddy, M. B. Nasr, B. E. A. Saleh, A. V. Sergienko and M. C. Teich, Phys. Rev. A  {\bf65}, 053817 (2002).
\bibitem{Strekalov1998} D. V. Strekalov, T. B. Pittman and Y. H. Shih, Phys. Rev. A {\bf57}, 567–570 (1998).
\bibitem{Nasr2003} M. B. Nasr, B. E. A. Saleh, A. V. Sergienko and M. C. Teich, Phys. Rev. Lett. {\bf91}, 083601 (2003).
\bibitem{Banaszek2007} K. Banaszek, A. S. Radunsky and I. A. Walmsley, Opt. Commun. {\bf269}, 152–155 (2007).
\end{thebibliography}
\end{document}